\documentclass[a4paper,12pt,english,german]{article}

\usepackage{graphicx}
\usepackage{amsmath}
\usepackage{mathtools}
\usepackage{graphicx}
\usepackage{hyperref}
\usepackage{amsmath}
\usepackage{amsthm}

\usepackage{mathtools}
\usepackage{amssymb}

\begin{document}

\centerline{\bf Macroscopic quantum distinguishablity, identity and irreversibility} 

\centerline{\bf in the Local Time Scheme}

\centerline{J. Jekni\' c-Dugi\' c$^{1}$, M. Arsenijevi\' c$^{2}$ and M. Dugi\' c$^{2}$}

\centerline{$^{1}$Department of Physics, Faculty of Sciences and Mathematics, 18000 Ni\v s, Serbia}

\centerline{$^{2}$Department of Physics, Faculty of Science, 34000 Kragujevac,
Serbia}

{\bf Abstract} The basic characteristics of the classical many-particle (''macroscopic'') systems are notoriously hard to reproduce in quantum theory. In this paper we show that this is not the case for certain many-particle systems within the recently introduced theory of emergent local quantum times, the so-called, Local Time Scheme. For an  isolated many-particle system consisting of large number of (approximately) isolated subsystems, distinguishability and individuality  can be naturally and straightforwardly obtained. In effect, a single such many-particle system quickly evolves between the mutually approximately orthogonal states thus setting a trajectory in the state space that is not shared with any other such individual many-particle system. Irreversibility of such dynamical processes is justified for the individual systems but not for the statistical ensembles of such systems. As an application, we derive a genuine nonexponential law for decay of unstable systems, in accordance with some theoretical expectations. This, classically plausible, picture calls for detailed analysis  regarding the relativistic causality and cosmological contexts.

\section{Introduction}

\noindent Practical indistinguishability or a lack of individuality are typically regarded to be a consequence of  limited information (i.e. ignorance) about the object (about the system's state).
Thence the methodological basis of statistical methods used as a description of the many-particle physical systems.

Despite almost one hundred years of scientific progress, those rules largely remain an elusive goal of quantum theory.
There is a large body of literature devoted to, or just tackling those topics, which are typically found within the ''quantum interpretations''
\cite{Goldstajn,MWI,QuBism,Ruth,Omnes}  or alternative quantum theories \cite{GRW} though with a limited success and a remarkable absence of the general basis.

In this paper we perform a technical analysis while leaving the interpretational details out of the considerations. Our approach to the topics of interest stems from a recently developed alternative approach to the concept of time within the quantum-mechanical context, the so-called Local Time Scheme (LTS) \cite{PRSA1,PRSA2}. In LTS, physical time appears as an  {\it emergent dynamical}  property of a {\it single} (''local''), isolated quantum system (which is, at least approximately, described by the unitary Schr\" odinger dynamics). Hence Time is not any of the fundamental physical notions but a consequence of a minimalist reasoning \cite{Kitada,MiPLA} about the unitary Schr\" odinger dynamics of quantum systems. Local Time carries a fundamental indeterminacy, meaning that a proper statistical ensemble of the  isolated systems assigns a definite local time  to every single element of the ensemble. So far, the basics of LTS have been developed in the context of the statistical-ensemble considerations of the structure-less or bipartitions of quantum systems. However, this framework limits the opportunity to tackle the {\it single} elements of an ensemble of {\it many-particle} systems and investigate the topics of macroscopic characetristics as outlined by the above rules.

In order to investigate those topics within LTS, in this paper we perform an analysis of the basic rules of LTS in the context of the individual, single many-particle isolated quantum systems. We apply the basic rules of LTS to a single many-particle isolated system and investigate dynamical emergence and change of the local time distribution for the subsystems of the total isolated system. On this basis, we are able naturally to obtain some quantum-mechanically substantial marks of both macroscopic distinguishability (MD) and macroscopic individuality (MI), and also of the Macroscopic Irreversibility (MIr), which is recognized as another characteristic trait of the classical macroscopic physical systems \cite{Zeh}. Within LTS, appearance of MD, MI and MIr is rather straightforward and reserved only for a many-particle isolated quantum system, which consists of a numerous set of mutually non-interacting isolated subsystems. Our analysis is technical in that we use some quantitative criteria for all of the desired macroscopic-system characteristics, MD, MI and MIr.

Assumption of noninteracting systems in a many-particle setting is of course an approximation, yet of general importance. Without this approximation (i.e. an idealization), the basic concepts of ''free'' (or even isolated) particles or fields cannot be introduced. The unitary Schr\" odinger dynamics could not be even formulated while some important cosmological models \cite{Hartle}  would become practically meaningless without the concept of ''at least approximate'' non-interacting (approximately isolated) systems. Isolated quantum system is subject of the unitary Schr\" odinger dynamics that almost invariably applies to the quantum-measurement \cite{zurek,Slosi} and decoherence theory \cite{joos} and models, and also presenting a basic assumption of the theory of open quantum systems \cite{breuer,Rivas}. As we show below, and thus set a basis of our considerations in this paper, the concept of isolated subsystems can be given a technical form within LTS.

In Section 2 the basics of LTS are presented and then extended in  details sufficient for the considerations performed in the rest of this paper. In Section 3, a classical-like dynamics for a single many-particle system is technically derived for the systems consisting of large number of isolated subsystems, some of which can be many-particle themselves. Section 4 presents details regarding different notions of (ir)reversibility. Section 5 is an overview of the results obtained in the previous sections, with an emphasis on the lack of distinguishability and individuality of quantum subsystems.
In Section 6, we analyze the unstable-system decay as an illustration of our general considerations and obtain non-exponential law in accordance with some general theoretical expectations. Section 7 is discussion and we conclude in Section 8.

\section{Many-particle systems}\label{sec:2}

\subsection{Outlines of LTS}\label{sec:21}

In a minimalist thinking about the Schr\" odinger unitary dynamics, $\hat U(t)=\exp(-\imath\hat H t/\hbar)$ where $\hat H$ is the isolated-system's Hamiltonian, a concept of Local Time comes first \cite{MiPLA}. The parameter $t$ defines a family of unitary operators while the Hamiltonian generating dynamics according to the unitary law introduces {\it uniqueness relation} \cite{MiPLA}:

\smallskip

\centerline {(R1) {\it One Hamiltonian} $\mathcal{H} \longleftrightarrow$ {\it One local time for some large values of the parameter} $t$.}

\smallskip

\noindent Every value of  the parameter $t$ is commonly assumed to represent a time-instant in the standard quantum theory with unique time (in a fixed, classical reference frame). In LTS, for small values of $t$, the local time is not well defined. Only for some large value $t_{\circ}$ of the parameter $t$, an individual-system's local time may be introduced \cite{PRSA1}. However,
this value is {\it not uniquely} determined \cite{PRSA1}. Then, instead of the fundamental quantum-mechanical law for all the individual systems equally prepared in an initial state $\vert\Psi(0)\rangle$,
\begin{equation}\label{eqn:1}
\vert \Psi(t_{\circ})\rangle =  \hat U(t_{\circ}) \vert \Psi(0)\rangle
\end{equation}

\noindent indeterminacy of the ''sufficiently large'' value of the parameter $t_{\circ}$ introduces the alternative fundamental equation of dynamical evolution in LTS \cite{PRSA1}:
\begin{equation}\label{eqn:2}
\hat \sigma(t_{\circ}) = \int_{t_{\circ}-\Delta t}^{t_{\circ} + \Delta
t} dt \rho(t) \vert \Psi(t)\rangle\langle\Psi(t)\vert =
\int_{t_{\circ}-\Delta t}^{t_{\circ} + \Delta t} dt \rho(t) \hat U(t)
\vert \Psi(0)\rangle\langle\Psi(0)\vert \hat U^{\dag}(t),
\end{equation}

\noindent where $\rho(t)$ is chosen without loss of generality as a Gaussian probability distribution sharply peaked around $t_{\circ}$, and the interval of indeterminacy $\Delta t$ is constrained by $\Delta t<\tau_{min}=\max\{\pi\hbar/2\Delta\hat H,$  $\pi\hbar/2(\langle\hat H\rangle)-E_g\}$ \cite{PRSA1}; $\Delta\hat H$ and $E_g$ represent the standard deviation and the ground value of the isolated system Hamiltonian $\hat H$; extension of equation (\ref{eqn:2}) to mixed states is straightforward. Equation (\ref{eqn:2}) fulfils the energy conservation condition, $\langle\hat H(t)\rangle=tr\hat H\hat\sigma(t_{\circ})=const$ and provides
that the energy eigenprojectors are invariants of the map. Hence the thermal state $\hat\rho_{th}$ (of the microcanonical or canonical or the grand canonical ensemble) is invariant of the LTS dynamical map, i.e., if the initial state is $\hat\rho_{th}$, the ''final'' state in equation (\ref{eqn:2}) $\hat\sigma(t_{\circ})=\hat\rho_{th}$.

The mixed state $\hat\sigma(t_{\circ})$ introduces a statistical ensemble of systems described by the same Hilbert state space and the Hamiltonian (as well as the underlying degrees of freedom). The point strongly to be stressed is that the ''mixture'' presented by equation (\ref{eqn:2}) is {\it proper} \cite{Despa}: every single element of the ensemble is subject of {\it its own}, uniquely determined Local Time, i.e. by its own time-instant $t_i\equiv t_{\circ}+\delta t_i$, where $\delta t_i\in[t_{\circ}-\Delta t,t_{\circ}+\Delta t]$. Therefore, not only that different Hamiltonians (different kinds of physical systems), but also the different systems of the same physical kind have different ''time axes'':

\smallskip

\centerline{(R2) {\it One individual (isolated) system} $\longleftrightarrow$ {\it One Local Time}.}

\smallskip

\noindent If (for sufficiently large $t_{\circ}$) an individual system is described by the unitary dynamics $\vert\Psi(t_i)\rangle=\exp(-\imath\hat H t_i/\hbar)\vert\Psi(0)\rangle$, the time instant $t_i$ can be recognized as its own time instant, which is very close to $ t_{\circ}$, while the indeterminacy-interval $\Delta t$ is chosen as to $\vert\langle\Psi(t_{\circ}+\Delta t)\vert\Psi(t_{\circ}-\Delta t)\rangle\vert > 0$ \cite{PRSA1}. Albeit mutually non-orthogonal, the states $\vert\Psi(t_i)\rangle$ demonstrate that individual systems in the ensemble described by equation (\ref{eqn:2}) are of ''different age''.

The structure-less and the systems with few degrees of freedom are quantitatively well described by the time instant $t_{\circ}$ thus justifying the use of the universal time in the standard quantum theory. However, for the many-particle systems, nonzero (albeit very small) $\Delta t$ leads to the non-negligible role and implications of Local Time \cite{PRSA1,PRSA2}. It cannot be overstated: arbitrariness of the local time instant $t_i=t_{\circ}+\delta t_i$ is {\it not} a consequence of any stochastic process. It is just a consequence of the impossibility uniquely to determine time for an individual quantum system.
Hence a local time as a hidden classical parameter of the system's dynamics.

If an isolated system $S$ consists of a pair of mutually noninteracting isolated subsystems $S_1$ and $S_2$, the rules (R1) and (R2) imply existence of local times for both subsystems, independently of each other. Then the total system $S$ is {\it not} determined by its own local time. However, if the subsystems mutually strongly interact, they form a total system $S$ that is described by its own local time, which is shared also by both subsystems $S_1$ and $S_2$. Then the common time for $S$, $S_1$ and $S_2$ is determined by the well-defined $t=0$ \cite{PRSA2}, which appears as the initial value $t=0$ on the most rhs of equation (\ref{eqn:2}). That is, merging (by sufficiently strong interaction) of the two isolated systems (subsystems $S_1$ and $S_2$) introduces the ''start'' of the total system's ''life''.

 LTS naturally differentiates between the few- and
many- particle systems, routinely and technically-simply describes
quantum measurement, resolves the 'preferred pointer-basis
problem' for an ensemble of bipartitions of closed many-particle
systems and provides a plausible interpretation of the
Wheeler-DeWitt equation \cite{PRSA1}. Formally, the map implicit to equation (\ref{eqn:3}) is non-differentiable but effectively Markovian with a coarse graining of the local time in accordance to the so-called Born approximation \cite{PRSA2}.
It is worth emphasizing: in LTS, every quantum system subject of a unique time is invariably subject of the standard quantum theory description.

\subsection{Dynamical emergence and redistribution of local times}\label{sec:22}

The mixed state $\hat\sigma(t_{\circ})$ in equation (\ref{eqn:2}) describes a ''proper mixture'' \cite{Despa}, i.e. a statistical ensemble in the Gibbsian sense of the standard interpretation of the concept of probability \cite{Hrenikov}: every single element of the ensemble goes through a unitary dynamics with the ''final'' local time instant, which is not shared with any other element of the ensemble (except in a coarse grained picture). A change of the local time may occur if the isolated system sufficiently strongly interacts with some other system (''merging'' the systems) thus forming a new total isolated system that can be subject of its own local time \cite{PRSA1,PRSA2}.

On the other hand, the rules (R1) and (R2) also describe the opposite process of splitting of an isolated system into mutually independent subsystems. If the total system defined by a local time can be dynamically described by a pair of mutually noninteracting and (approximately) isolated subsystems, then the process of splitting (as opposite to the above described process of  merging) yields a pair of isolated systems which have their own local times with well-defined $t=0$ for both subsystems.
However, there arise certain open questions of the basic physical importance. First, if Time is not a basic concept, what might be the primitive(s) of LTS in an attempt of axiomatization? Second, what can be told about the global picture where there is only one isolated system--the ''universe''?  Third, what might be description of a dynamical change of structure (separation into subsystems) and redistribution of local times in a many-particle setting?
Fourth, what are the possible descriptions and characteristics of the single elements of a statistical ensemble, particularly regarding the many-particle systems?

In answer to the first two questions, as a primitive [the very basic concept] of LTS, we introduce {\it dynamics} for an exactly isolated quantum system. ''Dynamics'' is simply a chain of the state transitions (state mappings):
\begin{equation}\label{eqn:3}
\dots \to \vert\Psi_{\circ}\rangle \to \dots \to \vert\Psi_{i}\rangle\to \vert\Psi_{i+1}\rangle\to \dots \to\vert\Psi_n\rangle\to\dots
\end{equation}

\noindent In line with the standard cosmological assumption, we deal exclusively with the pure quantum states of the ''universe''. The arrows in equation (\ref{eqn:3}) are assumed to be unitary. Equation (\ref{eqn:3}) is a discretized trajectory in the Hilbert state space of the system. The subsystems' states are obtained via the standard tracing out operation for every state in the chain (\ref{eqn:3}). If every subsystem (degrees of freedom) is in sufficiently strong interaction with at least one subsystem, they all share the same local time, which is also the local time for the total composite system. Then every single element of the ensemble (\ref{eqn:3}) is invariably subject of the standard quantum theory.

In regard of the above third question, we emphasize that the total system (the ''universe'') is described by some appropriate state space (elements of which are chained in equation (\ref{eqn:3})) and the system's Hamiltonian:

\begin{equation}\label{eqn:4}
\hat H = \sum_i \hat T_i +\sum_{i,j(\neq i)} \hat H_{ij}
\end{equation}

\noindent where $\hat T$ stands for the kinetic energy and $\hat  H_{ij}$ for the pair interaction between the subsystems $i$ and $j$ (for some details see \cite{MiPLA}).
The dynamical chain (\ref{eqn:3}) is governed by the unitary dynamics generated by the Hamiltonian (\ref{eqn:4}) and, in the standard theory, parametrized by unique time, i.e. $\hat U(t)=\exp(-\imath \hat Ht/\hbar)$. However, the rules (R1) and (R2) impose a completely new picture of the isolated system dynamics as we are going to present.

Every term in equation (\ref{eqn:4}) is with the uniquely defined mean value for every state appearing in the chain (\ref{eqn:3}). Then the negligible mean values of some of the interaction terms $\langle\hat H_{ij}\rangle$ in equation (\ref{eqn:4}) may define a split of the total system into (approximately) isolated subsystems, for some state $\vert\Psi_i\rangle$ in  equation (\ref{eqn:3}). In general, a split valid for some state $\vert\Psi_i\rangle$ may not apply to some other state $\vert\Psi_j\rangle$. For every such split of the total system, the LTS rules (R1) and (R2) should be {\it separately applied to every isolated subsystem} thus possibly introducing a new distribution of local times. If some another state in the chain (\ref{eqn:3}) emphasizes another split and hence another local time distribution, we can recognize dynamical emergence and redistribution of local times of the ''universe''. This sets a basis for tackling the above fourth question, which will be answered in the rest of this section.

Going back to the bipartite-system example of Section 2.1, in some state $\vert\Psi_i\rangle$ of the total system, a pair of isolated subsystems is described by the Hamiltonian $\hat H\approx \hat H_1+\hat H_2$ and therefore, according to the rules (R1) and (R2),  by a local-time distribution $\{\mathcal{T}_1,\mathcal{T}_2\}$. This scenario is described by the unitary dynamics $\exp(-\imath\hat H_1t_1/\hbar)\otimes\exp(-\imath\hat H_2 t_2/\hbar)\vert \Psi_i\rangle$. Then merging the subsystems valid {\it for} some state $\vert\Psi_{i+1}\rangle$, such that $\vert\Psi_i\rangle\to\vert\Psi_{i+1}\rangle$, is described by emergence of the common local time $\mathcal{T}$, which is dynamically induced by a sufficiently strong interaction $\hat H_{12}$ and presented by the unitary dynamics $\exp(-\imath(\hat H_1+\hat H_2+\hat H_{12})t/\hbar)$. More schematically, the dynamical transition $\vert\Psi_i\rangle\to\vert\Psi_{i+1}\rangle$ of the subsystems' merging is also presented:
\begin{eqnarray}\label{eqn:5}
\nonumber
    &\{\mathcal{T}_1,\mathcal{T}_2\} \to \mathcal{T} \\
    &\hat U^{split}\equiv e^{-\imath\hat H_1 t_1/\hbar}\otimes e^{-\imath\hat H_2t_2/\hbar} \to \hat U^{merge}\equiv e^{\imath(\hat H_1+\hat H_2+\hat H_{12})t/\hbar} \\\nonumber
\end{eqnarray}

\noindent as a local-times emergence and redistribution. If any of the subsystems is also composite, existence of a unique local time for that subsystem means (as said above) that all of its subsystems are also subject of the same local time. Now the inverse to equation (\ref{eqn:5}) describes a split of the total system $S$ into subsystems and a local-times redistribution.

It is essential to note that, while the change in the unitary dynamics factorization  in equation (\ref{eqn:5}) is assumed to be {\it deterministic}, the choice of the local time instants is {\it non-deterministic}, ''accidental'', and not induced by any external mechanism. In simplified terms, for the many-particle settings, LTS can be imagined as the standard-theory restructuring (merging/splitting) within the total system endowed by the accidental choice of local time for every individual system that is (approximately) isolated from the rest of the total system.

Hence the {\it operational algorithm} for determining a local-time distribution of a bipartite isolated system for every {\it single} state $\vert\Psi_{\alpha}\rangle$ picked from the chain of equation (\ref{eqn:3}): (i) calculate the mean values of all the interaction terms $\langle\Psi_{\alpha}\vert\hat H_{pq}\vert\Psi_{\alpha}\rangle$; (ii) on the basis of the first step detect if there are approximately isolated subsystems by checking which unitary operator, $\hat U^{split}$ or $\hat U^{merge}$, gives a {\it better description} of the total system's dynamics {\it for the  state} $\vert\Psi_{\alpha}\rangle$; (iii) in accordance with the answer to the step (ii), either the $\{\mathcal{T}_1,\mathcal{T}_2\}$ or $\mathcal{T}$ local-time-distribution applies for the state $\vert\Psi_{\alpha}\rangle$, while large values are assumed for all the parameters, $t,t_i,i=1,2$; (iv) the procedure (i)-(iii) should be repeated for every state known from the chain (\ref{eqn:3}). Extension of the procedure to a many-particle system is straightforward.

Consider a four-particle ''universe'':
\begin{eqnarray}\label{eqn:Scheme}
\nonumber
    &(1+2)\quad\cup \quad (3+4) \longrightarrow 1 \quad\cup \quad (2+3)\quad\cup\quad 4 \\\nonumber
    &\vert\Psi_{\alpha}\rangle  \longrightarrow \vert\Psi_{\beta}\rangle \\
    &\hat U_{12}(t_{12})\otimes\hat U_{34}(t_{34}) \longrightarrow \hat U_{1}(t_1)\otimes\hat U_{23}(t_{23})\otimes\hat U_{4}(t_4) \\\nonumber
\end{eqnarray}

Equation (\ref{eqn:Scheme}) illustrates a spontaneous, i.e., non-externally induced, restructuring of a composite system into (approximately) isolated subsystems in accordance with some state transition $\vert\Psi_{\alpha}\rangle\to\vert\Psi_{\beta}\rangle$ in equation (\ref{eqn:3}). Restructuring of the total system consists of the split of the initial bipartite subsystems and merging of the subsystems $2$ and $3$. The transition brings about emergence of the new local times and the local-times redistribution in the composite system. On the left: bipartite structure is described by the local time distribution $\{\mathcal{T}_{12},\mathcal{T}_{34}\}$. On the right of equation (\ref{eqn:Scheme}), a tripartite system with a novel time-distribution $\{\mathcal{T}_{1},\mathcal{T}_{23},\mathcal{T}_{4}\}$.
Dynamics of the total system is $\vert\Psi_{\beta}\rangle\equiv\vert\Psi(t_{12},t_{34})\rangle =\hat U_{12}(t_{12})\otimes\hat U_{34}(t_{34})\vert\Psi_{\alpha}\rangle$, where $\hat U_{ij}=\exp(-\imath (\hat H_i+\hat H_j+\hat H_{ij})t_{ij}/\hbar)$ with the local time $t_{ij}$ chosen for a single $S_{ij}$ system in accordance with the rules (R1) and (R2). For the general-form state $\vert\Psi_{\alpha}\rangle= \sum_{i,j,k,l}c_{ijkl}\vert i\rangle_1\vert j\rangle_2\vert k\rangle_3\vert l\rangle_4$ (we omit the symbol of tensor product), the state dynamics is of the form of:
\begin{equation}\label{eqn:7}
\vert\Psi_{\beta}\rangle\equiv\vert\Psi(t_{12},t_{34})\rangle = \sum_{i,j,k,l} c_{ijkl} \left(\hat U_{12}(t_{12}) \vert i\rangle_1 \vert j\rangle_2\right) \left(\hat U_{34}(t_{34})\vert k\rangle_3\vert l\rangle_4 \right).
\end{equation}

\noindent In full analogy follows the dynamics for the rhs of equation (\ref{eqn:Scheme}), where appears the total-system unitary operator $\hat U(t_{1})\otimes\hat U_{23}(t_{23})\otimes\hat U_{4}(t_{4})$.

Therefore, dynamical emergence and redistribution of local times in an isolated composite system is completely described by a chain of the tensor re-factorization of the total system's unitary operator in equation (\ref{eqn:Scheme}).
Extension to an arbitrary multi-particle system is straightforward, always with the rule of {\it dynamical emergence and possibly redistribution of local times}, e.g., instead of the second row in equation (\ref{eqn:Scheme}), we can shortly write:
\begin{equation}\label{eqn:8}
\vert\Psi(t_{12},t_{34})\rangle \to \vert\Psi(t_1,t_{23},t_4)\rangle.
\end{equation}

\subsection{Continuous trajectory in the state space}\label{sec:23}

Dynamical chain presented by equation (\ref{eqn:3}) is in the manner of the general quantum-operation formalism \cite{Kraus} that is virtually universally used in the quantum information and computation science \cite{Nilsen}.

In the standard quantum theory, the parameter $t$ is simply a time instant of the universal time axis defined and measured in a classical reference frame. Then the dynamical chain (\ref{eqn:3}) is just a coarse-grained description of a trajectory in the Hilbert state space of the total system. Symbolically, every trajectory can be presented by $\vert\Psi(t)\rangle$, where the choice of a specific value $t_i$ of the parameter $t$ corresponds to a state $\vert\Psi_i\rangle$ in equation (\ref{eqn:3}).
Continuous values of the local-time parameters as well as their coarse-grained pictures straightforwardly extend the one-time picture and the related formulas, such as the one presented by equation (\ref{eqn:Scheme}).

Local Time redistribution illustrated by equation (\ref{eqn:Scheme}) assumes that a given tensor-factorization of the total unitary operator {\it better describes dynamics} for the related ''initial'' state $\vert\Psi_{\alpha}\rangle$. But what if, for some special state $\vert\Psi_{\ast}\rangle$, the two unitary operators give {\it exactly} the same result? While such situations may be rare or even pathological, they should be considered. For example, equation
\begin{equation}\label{eqn:9}
\hat U_{12}(t_{12})\otimes\hat U_{34}(t_{34})\vert\Psi_{\ast}\rangle = \hat U(t_{1})\otimes\hat U_{23}(t_{23})\otimes\hat U_{4}(t_{4})\vert\Psi_{\ast}\rangle
\end{equation}

\noindent clearly implies undefined local times. Certainly, this indeterminacy can be avoided in the coarse-grained description of equation (\ref{eqn:3}).

We can summarize different indeterminacies linked with the concept of Local Time: (a) For small values of the $t$-parameter, local time is not yet established; (b) A local time is a hidden parameter of the ''local'' system's dynamics; (c) a composite system consisting of a set of isolated subsystems cannot be assigned a unique time; (d) there may be some exceptional case like the one presented by equation (\ref{eqn:9}), for which even the parameter $t$ is not uniquely defined. However, as demonstrated in Ref. \cite{PRSA1}, formation of a local time for many-particle systems  occurs in the rather short intervals of the parameter $t$--of the order of the time intervals known as the ''decoherence time'' in the decoherence/open systems theory \cite{zurek,Slosi,joos,breuer}. Hence, for most of the {\it practical} purposes, description of the many-particle systems may be based on the {\it approximate well-defined local times distribution} and hence a continuous trajectory in the state space of the total system (of the ''universe'') that re-writes equation (\ref{eqn:7}) of the form of
\begin{equation}\label{eqn:new}
\vert\Psi(t_1,t_{23},t_{4})\rangle = \hat U(t_1,t_{23},t_{4})\hat U(t_{12},t_{34})\vert\Psi_{\alpha}\rangle,
\end{equation}

\noindent with obvious extension for every part of the dynamical chain (\ref{eqn:3}) and for arbitrary many-particle system.

\section{Macroscopic states orthogonality}\label{sec:3}

The standard quantum mechanics is insensitive to the number of particles in the system. The rules for introducing the degrees of freedom \cite{carcasi,LAP}, building the Hilbert state space and constructing the related observables are practically universal. Thence no a priori rules for distinguishing between the many-particle systems.

Of course, phenomenology teaches us that not all the quantum particles of the same kind are subject of the identity rules. For example, electrons belonging to independent atoms are not subject of the Pauli exclusion principle. Rather, only electrons belonging to one of the atoms should be treated as mutually identical, i.e. a subject of the Pauli principle--as it is known from construction of  the Mendeleev's periodic table of chemical elements. Interestingly, which electrons should be regarded identical is an easy question to answer within LTS: all electrons that are subject of the same local time should be subject of the Pauli exclusion principle.

In classical physics, distinguishing between the objects of the same physico-chemical kind can be performed via distinguishing between the states (and trajectories in the state space) of the objects. No matter how close in the state space, every pair of classical-system states can be distinguished--it is said that the classical states are mutually {\it orthogonal} \cite{Nilsen}. However, quantum systems are plagued by  indistinguishability of states--the states are typically nonorthogonal to each other.  This nonorthogonality is at the root of all of the basic quantum information/computation protocols/algorithms \cite{Nilsen}. That is, the states non-orthogonality is a ''resource'' for achieving some information/computation goals, which stay out of the reach in the classical settings \cite{Nilsen}. Quantum decoherence devastates the quantum ''resources'' \cite{Slosi,joos,breuer,Nilsen}. In effect, an experimenter can perform practically only on a set of the mutually orthogonal states, just like in classical physics--e.g. a ''decohered'' quantum computer would be able to perform only classical computation \cite{Nilsen}. Therefore, it is often said that quantum decoherence makes a quantum system effectively classical by providing only (approximately) orthogonal quantum states. However, quantum decoherence {\it requires environment}, i.e. an outside system (degrees of freedom) in interaction with the system (which is open system) of interest \cite{Slosi,joos,breuer,Rivas,Nilsen}.

Interestingly, as we are going to demonstrate, approximate orthogonality of states of some {\it isolated} many-particle systems is a natural and straightforward consequence of the basic rules of LTS.

\subsection{Orthogonal state transitions}\label{sec:31}

The typical quantum states are not the Hamiltonian eigenstates. Such an state can be unitarily mapped into a ''final state'' that is (approximately) orthogonal to the ''initial'' one. That is, unitary dynamics $\hat U(t)=e^{-\imath t\hat H/\hbar}$ can transform a state $\vert\Psi\rangle$ into approximately orthogonal state $e^{-\imath\hat H/\hbar}\vert\Psi\rangle$ of the same, individual (single) system. The time bound (a minimum time interval needed for the orthogonal state transition) can be estimated larger than $\tau=\max\{h/4(\langle\hat H\rangle-E_{g}),h/4\Delta \hat H\}$ \cite{Margolus}, where $\hat H$ is the system Hamiltonian, $\langle\hat H\rangle - E_g$ its average value above the ground energy $E_g$ and $\Delta\hat H$ is the energy standard deviation for the initial state of the system.

Typically, we expect low energy for the many-particle systems thus reducing the time bound to the first term, $\tau=h/4(\langle\hat H\rangle-E_g)$. If the Hamiltonian $\hat H$ describes a system of interacting particles that are subject of the same (local) time, then $\langle\hat H\rangle=\langle \hat h\rangle+\langle \hat h_{int}\rangle$ , where $\hat h$ stands for the total self-Hamiltonians, and $\hat h_{int}$ for the set of the interaction terms of the total Hamiltonian $\hat H$. In such case, the standard quantum mechanical procedure applies, even if the constituting subsystems are many-particle themselves.

In LTS there are local times for non-interacting subsystems, every subsystem described by its own Hamiltonian $\hat H_i$ and the related local time $t_i$. If the spectral forms of the ''local'' Hamiltonians (discrete set of energies) are of the form, $\hat H_i=\sum_p E_{ip}\hat P_{ip}$, the condition of the orthogonal state transition for a {\it single} composite system (subject of a local time $t_i$) reads
\begin{equation}\label{eqn:A}
\langle\Psi\vert \otimes_i e^{-\imath t_i\sum_p E_{ip}\hat P_{ip}/\hbar}\vert\Psi\rangle=0.
\end{equation}

Without loss of generality, consider a bipartite system. Substituting the spectral forms of the local Hamiltonians and the use of the inequality $\cos x\ge 1-2(x+\sin x)/\pi$, gives for the real part of the scalar product (\ref{eqn:A})
\begin{equation}\label{eqn:B}
\begin{split}
0=\Re\left[ \sum_{p,q} e^{-\imath(E_{1p}t_1+E_{2q}t_2)/\hbar}\langle\Psi\vert\hat P_{1p}\otimes\hat P_{2q}\vert\Psi\rangle\right]=\\
\sum_{p,q} \langle\Psi\vert \hat P_{1p}\otimes\hat P_{2q} \vert\Psi \rangle\cos((E_{1p}t_1+E_{2q}t_2)/\hbar)\equiv \sum_{p,q}c_{pq}\cos((E_{1p}t_1+E_{2q}t_2)/\hbar)\ge\\
\sum_{p,q} c_{pq} \left[
1-{2\over\pi\hbar}
\left(E_{1p}t_1+E_{2q}t_2+\hbar\sin\left({E_{1p}t_1+E_{2q}t_2\over\hbar}\right)  \right)
\right].
\end{split}
\end{equation}

\noindent Since we are interested in satisfied equality in (\ref{eqn:A}), we search for the bound for which $\sin\left({E_{1p}t_1+E_{2q}t_2\over\hbar}\right)=0$. On the use of $\sum_{p,q}c_{pq}=1$, $\sum_{p,q}c_{pq}E_{1p}=\langle\hat H_1\rangle$ and $\sum_{p,q}c_{pq}E_{2q}=\langle\hat H_2\rangle$, equation (\ref{eqn:B}) gives $\langle \hat H_1\rangle t_1 + \langle\hat H_2\rangle t_2\ge {h/ 4}$. We are interested in the maximum time needed for the state transition, the choice of $\tau=\max\{t_1,t_2\}$ finally gives the bound
\begin{equation}\label{eqn:C}
\tau = {h\over 4 (\langle\hat H_1\rangle - E_{1g}+\langle\hat H_2\rangle-E_{2g})},
\end{equation}

\noindent where we count the energy starting from the ground energies of both subsystems, $E_{ig},i=1,2$. Generalization is obvious:
\begin{equation}\label{eqn:D}
\tau = {h\over \sum_i(\langle \hat H_i\rangle-E_{ig})}.
\end{equation}

From equation (\ref{eqn:D}) we can learn that the time bound (the minimum time needed for the orthogonal state transition) $\tau $ for a single composite system consisting of (approximately) isolated systems is smaller that the time bound $\tau_i$ for every constituent subsystem, and also that for a (single, individual) many-particle system consisting of a large number $N$ of isolated subsystems, $\tau^{(N)}$, goes to zero: $\tau^{(1)}>\tau^{(2)}>\dots>\tau^{(N)}$. That is, the orthogonal state transitions for the isolated large many-particle systems [consisting of a large number of isolated subsystems] are rather fast compared to the transitions of the smaller many-particle systems. As long as there is not the subsystems merging or splitting, a coarse grained picture of the system dynamics can give (approximate) orthogonal state transitions--like for the classical systems, where every {\it individual} many-particle system goes through successive transitions between the  orthogonal states.

\subsection{Distinguishability}\label{sec:32}

 A quantum state $\vert\Psi\rangle$ is often interpreted as a description of a statistical ensemble, e.g. \cite{Nilsen}, where every element of the ensemble is in the same (pure) state $\vert\Psi\rangle$. This statement is in the root of the procedure of the quantum state {\it preparation}. However, the state (\ref{eqn:2}) is a {\it proper} mixture of different states. If one single element of the ensemble is in a state $\vert\Psi\rangle$, another single element of the ensemble is in some other state $\vert\Psi'\rangle$. As we are going to demonstrate, scalar products of such states, $\langle\Psi\vert\Psi'\rangle$, is modulo very small--approximate orthogonality of states for a {\it single} pair of  elements of the ensemble (\ref{eqn:2}).

Consider a four-body system described by the state on the rhs of equation (\ref{eqn:8}), $\vert\Psi(t_1,t_{23},t_4)\rangle$, i.e. by the local-time distribution $\{\mathcal{T}_{1},\mathcal{T}_{23},\mathcal{T}_4\}$. If we apply the general rules of LTS, Section 2, to every ''local system'' separately, we can learn that a pair of individual systems, which belong to the same statistical ensemble, if described by the same state are still subject of the different local-time distributions. That is, one system is in the state $\vert\Psi(t_1,t_{23},t_4)\rangle$, while the other is in some state $\vert\Psi(t'_1,t'_{23},t'_4)\rangle$ and $t_{ij}\neq t'_{ij}\in[t_{\circ}-\Delta t_{ij},t_{\circ}-\Delta t_{ij}]$. Intuitively, in every pair of subsystems, one subsystem is ''older'' than the other one.

Even if the two composite systems have exactly the same initial $t=0$, they are {\it different} states belonging to the {\it same} Hilbert space. Then
$\vert\langle \Psi(t_1,t_{23},t_{4})\vert \Psi(t'_1,t'_{23},t'_{4})\rangle\vert < 1$.

Consider the total system's state $\vert\Psi(t_1,t_{23},t_4)\rangle=\sum_{i,j,k} c_{ijk}\hat U_1(t_1)\vert i\rangle_1\otimes\hat U_{23}(t_{23})\vert j\rangle_{23}\otimes\hat U_4 (t_4)\vert k\rangle_4$ from equation (\ref{eqn:8}). Then the scalar product of  states of the two individual systems
\begin{equation}\label{eqn:10}
\begin{split}
S(\mathcal{T}_1,\mathcal{T}_{23},\mathcal{T}_4):=\langle \Psi(t_1,t_{23},t_{4})\vert \Psi(t'_1,t'_{23},t'_{4})\rangle=\\
\sum_{i,j,k,i',j',k'} c^{\ast}_{ijk} c_{i'j'k'} \langle i \vert \hat U_1^{\dag}(t_1)\hat U_1(t'_1) \vert i'\rangle
 \cdot \langle j \vert \hat U_{23}^{\dag}(t_{23})\hat U_{23}(t'_{23}) \vert j'\rangle \cdot \langle k \vert \hat U_4^{\dag}(t_4)\hat U_4(t'_4) \vert k'\rangle=\\
 \sum_{i,j,k,i',j',k'} c^{\ast}_{ijk} c_{i'j'k'} \langle i \vert \hat U_1(\delta t_1) \vert i'\rangle
 \cdot \langle j \vert \hat U_{23}(\delta t_{23}) \vert j'\rangle \cdot \langle k \vert \hat U_4(\delta t_4) \vert k'\rangle,
 \end{split}
\end{equation}

\noindent where $\delta t_{\alpha}\equiv t'_{\alpha}-t_{\alpha}, \alpha=1, 2-3,4$.

In the standard theory there is only one time, i.e. $\delta t_{\alpha}=0,\forall{\alpha}$, thus giving for equation (\ref{eqn:10}): $\sum_{i,j,k,i',j',k'} c^{\ast}_{ijk} c_{i'j'k'}\delta_{ii'}\delta_{jj'}\delta_{kk'} = 1$. The choice of the basis vectors $\vert p\rangle, p=i,j,k$ for the subsystems to be eigenvectors of the subsystems' Hamiltonians yields
\begin{equation}\label{eqn:11}
S(\mathcal{T}_1,\mathcal{T}_{23},\mathcal{T}_4)=\sum_{i,j,k} \vert c_{ijk}\vert^2  e^{-\imath(E_{1i}\delta t_1+E_{23j}\delta t_{23}+E_{4k}\delta t_{4})/\hbar}.
\end{equation}

\noindent The sum in equation (\ref{eqn:11}) is the sum of the two-dimensional vectors (in the complex plane). If all vectors are closely oriented, the norm of the resultant vector (of the sum in equation (\ref{eqn:11})) is close to one. However, if the vectors are symmetrically oriented, their sum equals zero.

Consider (as we have already assumed) a narrow low-energy shell for the total system of the finite dimension $N$, which can be very large, and the normalized states $\vert\psi\rangle=\sum_{i=1}^{\nu}c_i\vert \varphi_i\rangle$, $\nu\le N$ in the shell. Assume the same statistical weights for all such states and introduce the quantity $u:=\sum_{i=1}^{\nu}\vert c_i\vert^2$.  On the use of the average value $\langle u^n\rangle = (\nu(\nu+1)\dots(\nu+n-1) )/(N(N+1)\dots(N+n-1))$ \cite{vN29}, we directly obtain $\langle\vert c_i\vert^2\rangle=1/N$ and $\langle\vert c_i\vert^4\rangle=2/N(N+1)$ by choosing $\nu=1$ and $n=1$, and $\nu=1$ and $n=2$, respectively. So we obtain that, typically, $\langle\vert c_i\vert^2\rangle=1/N$ with the very narrow distribution for large $N$--the standard deviation $\Delta(\vert c_i\vert^2)=\sqrt{(N-1)/N^2(N+1)}$. Therefore, for the approximately symmetrically distributed vectors in equation (\ref{eqn:11})--which may be expected to be fulfilled for many-particle systems (the sum in the exponent of equation (\ref{eqn:11}) should be at least of the order of $2\pi$, even for very small $\delta t$s)--and typically $\vert c_i\vert^2\sim 1/N$, very small value of the sum in equation (\ref{eqn:11}) can be expected.
Furthermore, the fact that for the many-particle systems (already for some large molecules), the energy spectrum is rather dense, a continuous limit may be applied \cite{Kitel}. Then, instead of the sum in equation (\ref{eqn:11}), we obtain a general form:
\begin{equation}\label{eqn:12}
S(\mathcal{T}_I,\mathcal{T}_{II},\mathcal{T}_{III},...)=\int_0^r p(x) e^{-\imath x} dx,
\end{equation}

\noindent with the probability density $p(x)$.

For the positive dense spectrum (typical for the many-particle systems), we can estimate the rhs of equation (\ref{eqn:12}) to be much smaller than one, i.e. approximately to be equal to zero--approximate {\it states orthogonality}. To be specific, assume the equal-probability distribution, i.e. $p(x)=r^{-1},\forall{x}$, which yields $S(\mathcal{T}_I,\mathcal{T}_{II},\mathcal{T}_{III},...)=\sqrt{2(1-\cos r)/r^2}$, which takes small value of approximately $0.19$ already for the modest $r\sim 10\approx 3.2\pi$ . In Appendix A we give an example of  a pair of few-body qubit-systems for which equation (\ref{eqn:11}) also returns very small value.
Hence we can conclude, that the states of a pair of individual many-particle systems--which consist of a large number of (approximately) isolated subsystems--are approximately orthogonal--macroscopic state orthogonality, i.e. {\it macroscopic distinguishability} from the Local Time Scheme.

\subsection{Individuality}

Distinguishability of the same-kind systems is based on the states that constitute a common state space of the systems. That is, the states and the related trajectories in the state space may, in principle, be characteristic of different systems of interest. However, the local time distribution and the choice of the local time instants distinguish individual systems as their characteristic trait. Despite the fact that local time is a hidden parameter of a system's dynamics, it distinguishes the individual system from the similar such systems. As long as there is not internal merging or splitting of subsystems, an individual isolated system is characterized by a local-time distribution and the choice of the local times as presented by the operator tensor-factorizations in equation (\ref{eqn:Scheme}) and (\ref{eqn:9}). Hence we propose a quantitative criterion of individuality without any reference to a system's state,
\begin{equation}\label{eqn:a}
\mathcal{I}:={1\over d}tr \left(\hat U^{\dag}(t'_1,t'_2,\dots)\hat U(t_1,t_2,\dots)\right)
\end{equation}

\noindent while assuming the same local-time distribution and different values of the time instants for the two individual systems with $d$ standing for the dimension of the total system's Hilbert state space. Equation (\ref{eqn:a}) is for operators--no states are involved. For the single system, i.e. for $t'_i=t_i,\forall i$, the maximum value $\mathcal{I}=1$ is attained.

For a single unitary operator (we assume $\hbar=1$) $\hat U(t)=e^{-\imath\hat H t}$, $\mathcal{I}=\sum_p e^{-\imath E_p\delta t/2}g_p/d$, where $E_p$ is the Hamiltonian eigenstate and $g_p$ its degeneracy. For equation (\ref{eqn:a})
\begin{equation}\label{eqn:b}
\mathcal{I}={1\over d}\sum_{i,j,\dots} e^{-{\imath}(E_{1i}\delta t_1+E_{2j}\delta t_2+\dots)} g_{1i}g_{2j}\dots
\end{equation}

\noindent with $d={\Pi}_i d_i$, where $d_i$ is dimension of the $i$th ''local''  subsystem's Hilbert space,  $\sum_{\alpha}g_{i\alpha }=d_i, i=1,2,\dots$, and $\sum_{i,j,\dots}(g_{1i}g_{2j}\dots)=d$. Obviously, equation (\ref{eqn:b}) is of the same kind as equation (\ref{eqn:11}), and hence, for a many-particle system, of the form of equation (\ref{eqn:12}), which implies analogous conclusion: the larger the total system, the smaller value of $\mathcal{I}$ thus distinguishing individual characteristics of the total isolated system.

\section{Macroscopic irreversibility}\label{sec:4}

''Reversibility '' is a physical, not mathematical concept, which assumes the time inverse of the actual process. Formally, if a process is presented by a dynamical map $\Phi(t,0)$, reversibility requires invertibility, i.e. existence of the inverse map $\Phi^{-1}(t,0)$. This condition is always fulfilled in the standard, unitary quantum theory of unique time. However, for the open systems, existence of the inverse map, albeit physically typical \cite{Invert}, does not present a physically reasonable process \cite{Rivas}. Therefore, reversibility requires invertibility {\it and} physically reasonable inverse of the process.

On the other hand, irreversibility seen in the ''macroscopic world'' is often linked with the phenomenological dissipation and sometimes introduced as a ''cause'' of the phenomenological ''arrow of time'' \cite{Zeh}. Hence, in general, irreversibility is not just the opposite to reversibility. In this section we discuss some aspects of (ir)reversibility within the Local Time Scheme. Therefore, it is useful to recall (as stated below equation (\ref{eqn:5})): while the local time redistribution is deterministic, the choice of the values of local times  is not.

\subsection{Plain irreversibility}

Dynamical map implicit to equation (\ref{eqn:3}) is  non-invertible \cite{PRSA1,PRSA2}. However, bearing in mind the meaning of the proper mixture described by the state $\hat\sigma(t_{\circ})$, the {\it individual} systems' dynamics are invertible. To see this, just consider one single system out of the ensemble $\hat\sigma(t_{\circ})$, i.e. a single system evolution $\hat U(t_i)$, where $t_i\in[t_{\circ}-\Delta t,t_{\circ}+\Delta t]$ is a local time for that single element of the statistical ensemble. Of course, there exists the inverse $\hat U^{-1}(t_i)$ that would return the state $\vert\Psi(t_i)\rangle$ into the initial state $\vert\Psi(0)\rangle$. This observation applies  for the general case; e.g. applying $\hat U_{12}^{\dag}(t_{12})\otimes\hat U^{\dag}_{34}(t_{34})$ to the state in equation (\ref{eqn:7}) returns the initial state. However, there is a caveat to this observation.

The inverse unitary dynamics for the {\it single} system may apply as long as there is {\it not a structural change} in the composite system. That is, while the single unitary-operators  are deterministic (and hence invertible \cite{Invert}), the choice of the subsystems' local time is {\it accidental}, i.e. non-deterministic. To be specific, the (expected, probable) inverse to the unitary transition in equation (\ref{eqn:7}) reads
\begin{equation}\label{eqn:13}
\hat U(t_{1})\otimes\hat U_{23}(t_{23})\otimes\hat U_{4}(t_{4})\to\hat U_{12}(t'_{12})\otimes\hat U_{34}(t'_{34}),
\end{equation}

\noindent where $t_{12}\neq t'_{12}$ and $t_{34}\neq t'_{34}$--where non-primed $t$s appear in equation (\ref{eqn:7}).
That is, every single system returns to a state that evolves according to another local time distribution than the original one. Intuitively, a  single system would practically never ''choose'' the time axis it had started in the original direction of dynamical evolution in equation (\ref{eqn:3})--e.g. the inverse to the state transition in equation (\ref{eqn:8}) reads
\begin{equation}\label{eqn:14}
\vert\Psi(t_1,t_{23},t_{4})\rangle\to\vert\Psi(t'_{12},t'_{34})\rangle\neq \vert\Psi(t_{12},t_{34})\rangle.
\end{equation}

\subsection{Stochastic reversibility}

In the previous section we used the standard logic of irreversibility: if the input of the direct, and the output of the inverse process may not coincide, the process is irreversible. In other words, whenever description of a system relies on the concept of probability, the standard logic implies irreversibility in the individual system of interest. However, in the context of the  stochastic processes, reversibility can be alternatively introduced in the statistical {\it ensemble} context\cite{Lynn}. Typically, ''stochastic process'' is linked with external influence, but this is not the case in some alternative quantum theories, e.g., in \cite{GRW} and in LTS \cite{PRSA1,PRSA2}.

For the classical stochastic processes, a process is said to be reversible if the condition of detailed balance is fulfilled \cite{Lynn}. If  $P(x\to x')$ denotes joint probability of the transition $x\to x'$ and $P(x'\to x)$ is joint probability of the inverse transition for the states $x,x'$, the condition of detailed balance is fulfilled if and only if $P(x\to x')=P(x'\to x),\forall{x,x'}$. Then the relative entropy \cite{Nilsen,Lynn}
\begin{equation}\label{eqn:15}
I:=\sum_{x,x'} P(x\to x') \log{P(x\to x')\over P(x'\to x)}
\end{equation}

\noindent equals zero. If for at least one pair of states $P(x\to x')\neq P(x'\to x)$,  the quantity $I>0$, thus exhibiting that the process is irreversible \cite{Lynn}.

The logic behind the use of equation (\ref{eqn:15}) is fully applicable in the context of our considerations, that is, the state transitions in equations (\ref{eqn:8}) and (\ref{eqn:14}) are susceptible to quantification by the relative entropy $I$.
For every structure (split into isolated subsystems) of a single, isolated composite system, the choice of a value of a local time instant is non-deterministic, independent of the choice regarding any other local time for that structure,  and independent of any other structure of the total system. The choice of a local time, and hence of a set of local times, is purely classical. Hence
the structural changes as described by equations (\ref{eqn:8}) and (\ref{eqn:14}) have the effects similar to the classical stochastic processes, which would determine both the structure and the chosen values of the local time instants.

For a given structure of the total system, probability of a combination of the concrete values of local times gives the probability of the states in the set. Formally, probability for the concrete (discretized) values of the local time $p(t_1,t_{23},t_4)$ is the probability of the state $\vert\Psi(t_1,t_{23},t_4)\rangle$ in the (''coarse-grained'') ensemble of equation (\ref{eqn:2}). Then transition e.g. in equation (\ref{eqn:8}), $\vert\Psi(t_{12},t_{34})\rangle \to \vert\Psi(t_1,t_{23},t_4)\rangle$, describes {\it independent} choices of the respective local times thus giving rise to the transition  probability $P(\vert\Psi(t_{12},t_{34})\rangle \to \vert\Psi(t_1,t_{23},t_4)\rangle)=p(t_{12},t_{34})p(t_{1},t_{23},t_4)$. Since the inverse state transition is with the same probability, the relative entropy $I=0$. Thus we can conclude that the local time redistribution is {\it reversible} in the considered stochastic sense.

\section{Dynamics of a single isolated many-particle system}\label{sec:5}

From the previous sections, we can learn, that a single many-particle system experiences the rather fast orthogonal state-transitions and can be (approximately)  distinguished from any other such system. If presented in a discrete form of equation (\ref{eqn:3}), a single isolated system consisting of a large number of (approximately) isolated subsystems goes through a {\it series of the (approximately)  orthogonal state-transitions} thus providing a ''history'' of the form of equation (\ref{eqn:3}), which is {\it not shared with any other such many-particle system}. Substituting the quantum states $\vert\Psi_i\rangle$ in equation (\ref{eqn:3}) with the classical state-space (phase space) states $\mathfrak{s}_i=\{\vec r_{\alpha}^i,\vec p_{\alpha}^{i}, \alpha=1,2,\dots N\}$, we obtain a discretized form of the classical phase-space trajectories. In this sense, we say that LTS reproduces a classical-like dynamics for a single isolated many-particle system, which consists of a large number of isolated subsystems, such that (a) there is neither internal splitting/merging of the subsystems (b) nor merging of the system with any other external system, (c) the system may be described by the pure-state dynamical chain of equation (\ref{eqn:3}) and (d) a ''local'' system carries individual characteristics as quantified by equation (\ref{eqn:a}). On the individual-system level, such systems dynamics is reversible. Irreversibility emerges for an individual system whenever the items (a) and/or (b) are not satisfied.

Even if the first two assumptions (a) and (b) are fulfilled, in general, the isolated subsystems of a large isolated whole typically do not fulfil the conditions (c) and (d).  The many-particle systems, even in the standard theory, are complex net of the internal subsystems correlations like in equation (\ref{eqn:7}). In quantum modeling, it is often assumed that initial state of a many-particle system, such as the open systems or the ''apparatus'' of quantum measurement, may be initially in a pure state. However, interaction with other subsystems quickly destroys the state purity and, typically, give rise to a mixed reduced-state of the subsystem \cite{ppsd}. Such states are ''improper mixtures'' \cite{Despa} for which individual system's dynamics is not generally known, albeit a subject of interpretations. Therefore, in general, the task of deriving macroscopic behavior of quantum subsystems, even if isolated in the sense of LTS, is not well posed a task. To this end, for completeness, in Appendix B, we briefly analyze the subsystems dynamics of the model of equation (\ref{eqn:7}). While the isolated subsystems' reduced states evolve unitarily, dynamics of nonisolated subsystems (parts of a larger isolated subsystems) are found in a diagonal form, e.g. equation (\ref{eqn:B7}), that may be a result of the internal decoherence process.

\section{Nonexponential decay in Local Time Scheme}\label{sec:6}

State decay is a generic effect observed in unstable systems. Nevertheless, as recognized in an extensive bibliography, it's quantum mechanical theoretical basis as well as direct experimental observation remain a challenge  even in the simplified scenarios. In practice, and particularly in nuclear physics, decays are typically described by an exponential law that is recognized as a result of ''somewhat delicate approximations'' and hence cannot be regarded exact \cite{Merz,Kinezi,Peres}. Rather, it is believed that non-exponential decay law is generic \cite{Kinezi}, in the sense that for short time intervals \cite{Nature} and for the very long times \cite{Peshkin}, a non-exponential law applies, while the exponential law can be adopted for the intermediate (possibly long, but not very long) times \cite{Peres,Peshkin}. It is also assumed that at the initial time (time instant $t=0$), the decay rate should vanish \cite{Fonda,Grinland}. On the level of individual systems, decays are recognized as stochastic processes that can be described within the open systems theory \cite{Kinezi}. It is therefore interesting to investigate consequences of LTS in the context of the individual (approximately) isolated decaying systems--a process that is simply splitting of a composite system in LTS.

Probability $p(t)$ of survival of undecayed state of an individual system is typically described by a constant rate $\lambda>0$, so that $dp=-\lambda dt$ and the exponential law $p(t)=e^{-\lambda t}$ is obtained. For an individual system with a local time $t_1$, a minimal extension of the standard approach gives $dp(t_1)=-\lambda dt_1$. If expressed by a local time $t$ of a laboratory clock, which gauges dynamics of other systems,
\begin{equation}\label{eqn:16}
dp(t)=-\lambda\kappa(t)dt,\quad \kappa(t)>0,\forall{t>0},
\end{equation}

\noindent thus giving rise to a more general law:
\begin{equation}\label{eqn:17}
p(t)=e^{-\lambda\int_0^t \kappa(s)ds}
\end{equation}

\noindent that should be observed in a laboratory; we assume that $p(0)=1$.

{\it Prima facie}, $\kappa(t)$ should equal one. Actually, getting back to a definition of an individual local time, $t_1=t_{\circ}+\delta t_1$ ($\delta t_1\ll t_{\circ}$), the two local times (here of the laboratory clock, $t=t_{\circ}+\delta t$, and of the decaying system, $t_1$)  differ only by a constant--like in the time-translation operation--thus implying $\kappa=1,\forall{t}$. However, this is not the case.

 From equation (\ref{eqn:2}), we can read that different individual systems, even of the same kind, start from the same initial $t=0$. Bearing in mind that, in general,  the local times should acquire different values even if they had the same start of $t=0$,
 their respective time-rates cannot be equal--consequently, $\kappa(t)\neq const.$, as a {\it universal rule}.
 Furthermore, for very large $t_{\circ}$, the two times $t_1$ and $t$ become mutually indistinguishable as well as indistinguishable from  $t_{\circ}$--the ratio $t_1/t=(t_{\circ}+\delta t_1)/(t_{\circ}+\delta t)$ goes to $1$ as $t_{\circ}\to\infty$ and hence the function $\kappa(t)$ should saturate to $1$. Finally, integration of $dt_1=\kappa_1(t)dt$ should provide $t_1=t+\delta t_1$, such that $\delta t_1\ll t$. Independently of a concrete form of $\kappa_1(t)$, the decay law (\ref{eqn:17}) is not exactly exponential: $p(t)=e^{-\lambda(t+\delta t_1(t))}$.

It would be interesting to see if a choice of $\kappa_1(t)$ may give the expected behavior \cite{Peres}: for short time intervals to start as a quadratic function and for the long time intervals to ''end'' as an inverse power law, while for the intermediate times to be describable by the standard exponential law. As we show below, this can be achieved within the above distinguished requirements: $\kappa_1(t)$ should be such that (i) $t_1=t+\delta t_1, \delta t_1\ll t$; (ii) $\kappa_1(t)>0,\forall{t}$; (iii) $\lim_{t\to\infty}\kappa_1(t)=1$.

Below, we do not present mathematically rigorous results. Rather, we construct a case as a proof of principle for achieving the desired goal. The condition (iii) can be satisfied by the choice of $\kappa_1(t)=P_n(t)/Q_n(t)$, where appear the same-order polynomials of $t$, as long as they have the same constant $a_n$. In order to fulfil the condition (ii), we choose all the polynomial coefficients to be positive. Then the condition (i) strongly depends on the concrete choice of the two polynomials.

Let us consider $P_2(t)=at+bt^2$ and $Q_2(t)=1+pt+bt^2$; the parameters $a,b,p$ can be phenomenologically determined or chosen. Thus the integral in  equation (\ref{eqn:17}) reads
\begin{equation}\label{eqn:18}
t_1 = \int {P_2(t)\over Q_2(t)} dt = t+ {(p^2-ap-2b)\arctan\left({p+2bt\over\sqrt{4b-p^2}}\right)\over b\sqrt{4b-p^2}}+{(a-p)\log(1+pt+bt^2)\over 2b}\equiv t+ \tau_1(t).
\end{equation}

\noindent Since  arctangent is a bounded and logarithm slowly increasing function, a choice of large $b$ and $4b>p^2$ can give $\vert\tau_1(t)\vert\ll t$ even for small $t$. Hence the above item (i) is satisfied.

On the use of equation (\ref{eqn:18}),  equation (\ref{eqn:17}) returns for small $t$
\begin{equation}\label{eqn:19}
e^{-\lambda(t+\tau_1(t))}\approx C\left(1-{a\lambda\over 2}t^2 \right), C \equiv e^{\lambda\mu} = e^{ \lambda {(2b+ap-p^2)\arctan{\left(  {p\over\sqrt{4b-p^2}}\right)}  \over b\sqrt{4b-p^2}}  },
\end{equation}

\noindent which is a quadratic law expected \cite{Peres} for short times. For medium $t$, $t+\tau_1(t)\approx t$ thus yielding approximately exponential law $e^{-\lambda t}$, while for very large $t$, approach to zero can be well approximated by the inverse law of $1/t^2$--as typically expected \cite{Peres}.

Equation (\ref{eqn:19}) offers an interesting and important lesson. According to the above point (i), $\tau_1\ll t$, one may suppose that the small-$t$ behavior is linear, i.e. that $p(t)\approx 1-\lambda(t+\tau_1(t))\approx-\lambda t$, which is of the first order of approximation. However, nontrivial time dependence of $\tau_1(t)$, as for the chosen polynomials $P_n(t)$ and $Q_n(t)$, may cancel out the first-order term and give the quadratic form as in equation (\ref{eqn:19}). That is, despite small value of $\tau_1(t)$, the decay probability $p(t)$ may carry non-trivial dependence on the parameter $t$ measured by a laboratory macroscopic clock.

Now we can describe a decay of a mother system (e.g. atomic nucleus) $A$ to a daughter system $B$. To this end, we need a probability distribution instead of the probability density $\rho(t)$ appearing in equation (\ref{eqn:2}). We can coarse grain the interval $[t_{\circ}-\Delta t,t_{\circ}+\Delta t]$ to obtain a set of time instants (for large $t_{\circ}$), $\{t_1,t_2,\dots\}$. Every $t_i\in\{t_1,t_2,\dots\}\subset [t_{\circ}-\Delta t,t_{\circ}+\Delta t]$ corresponds to a pure subensemble $\mathcal{S}_i$ with a statistical weight $p_i=N_i/N$, where $N$ stands for the total number of  elements in the ensemble. Then, for the fixed local time $t_A$ there is the number $N_A$ of the mother systems and $N_B$ of the daughter systems obtained via the decay process. On the use of the standard, one-time, law $dN_B=-\lambda_AN_Bdt+ \lambda_BN_Bdt$, the use of (\ref{eqn:16}) gives the differential law in LTS
\begin{equation}\label{eqn:20}
{dN_B\over dt} = -\lambda_B\kappa_B(t)N_B(t)+\lambda_A\kappa_A(t)N_A(t).
\end{equation}

\noindent Bearing in mind that now equation (\ref{eqn:16}) takes the form of $dN_A(t)=-\lambda_A\kappa_A(t)N_A(t)dt$, by matrix exponentiation we can obtain the decay-process dynamical map  and finally the solutions
\begin{equation}\label{eqn:21}
\begin{split}
N_A(t)=N_A(0)e^{-\lambda_A\int_0^t\kappa_A(s)ds}=N_A(0)e^{-\lambda(t+\tau_A(t))}, \\
 N_B(t)={N_A(0)\lambda_A\int_0^t\kappa_A(s)ds\over \lambda_B\int_0^t\kappa_B(s)ds-\lambda_A\int_0^t\kappa_A(s)ds}\left(e^{-\lambda_A\int_0^t\kappa_A(s)ds}-e^{-\lambda_B\int_0^t\kappa_B(s)ds}\right)=\\
 {N_A(0)\lambda_A(t+\tau_A(t))\over \lambda_B(t+\tau_B(t))-\lambda_A(t+\tau_A(t))}\left(e^{-\lambda_A(t+\tau_A(t))}-e^{-\lambda_B(t+\tau_B(t))}\right),
\end{split}
\end{equation}

\noindent with the initial conditions $N_A(t=0)\equiv N_A(0)$ and $N_B(t=0)=0$. The standard one-time solutions \cite{Merz} easily follow from (\ref{eqn:21}) for $\kappa_A(t)=1=\kappa_B(t),\forall t$, i.e. for $\tau_i(t)=0,i=A,B,\forall{t}$, e.g.
\begin{equation}\label{eqn:22}
N_B(t)={N_A(0)\lambda_A\over \lambda_B-\lambda_A}\left(e^{-\lambda_At}-e^{-\lambda_Bt}\right).
\end{equation}

From (\ref{eqn:21}) straightforwardly follows a short-time law
\begin{equation}\label{eqn:23}
N_B(t)\approx{N_A(0)\lambda_A(\mu_A-a_At^2/2)\over \lambda_B(\mu_B-a_Bt^2/2)-\lambda_A(\mu_A-a_At^2/2)}\left[1-{1\over 2}\left(C_Aa_A\lambda_A-C_Ba_B\lambda_B\right)t^2 \right],
\end{equation}

\noindent where we used equations (\ref{eqn:18}) and ({\ref{eqn:19}}) for both subsystems $A$ and $B$, which are indices to the corresponding quantities in (\ref{eqn:23}). It is obvious that (\ref{eqn:23}) is in contrast to the short-time limit of (\ref{eqn:22}), which yields
\begin{equation}\label{eqn:24}
N_B(t)\approx -\lambda_A N_A(0) t.
\end{equation}

\section{Discussion}\label{sec:7}

Merging and splitting the systems is an intuitive picture present in the standard quantum and classical theory. Merging of systems is formation of  composite systems via ''condensation'' (or ''capture'' of quantum particles), while splitting the systems is simply a break into independent pieces (e.g. via ionization, decay or breaking the chemical bonds). Within LTS, this picture (intuitively related to the spatial distance of the  systems) is endowed by the rules (R1) and (R2). Nonexistence of a unique time for a composite system consisting of (approximately) isolated subsystems has been proposed \cite{PRSA1} as an interpretation of the Wheeler-DeWitt equation $\hat H\vert\Psi\rangle=0$. A ''local'' (closed) system consisting of mutually interacting subsystems is subject of the standard quantum theory for the system as a whole as well as for every of its subsystems.

In this paper, we {\it derive} appearance of the classical-like distinguishability, individuality and irreversibility of {\it individual many-particle} systems consisting of a large number of (approximately) isolated individual subsystems. As another implication of Local Time Scheme, we demonstrate a possibility to describe the unstable-system decay as typically expected, particularly within nuclear physics, e.g. \cite{Peres}.

As long as dynamical chain (\ref{eqn:3}) may be adopted as a basis of description of an (approximately) isolated system, the results presented in the previous sections can be invariably used {\it if } there is not a dynamical change in the system's structure (merging or splitting of subsystems) or its merging with any other external system. However, in the context of quantum subsystems, equation (\ref{eqn:3}) typically cannot be  used. Rather, ''improper'' mixtures are used to describe quantum subsystems \cite{Despa} thus not justifying a mixed-states-chain analogous to equation (\ref{eqn:3}). Full analysis of quantum subsystems in the LTS context is not a part of this paper.

In the standard  quantum theory of unique time, restructuring (separation into ''parts'') of a composite system may introduce new subsystems (degrees of freedom), some of which may be dynamically separated (non-interacting, i.e. uncoupled, via the variables separation). In this regard, typical are the system's ''collective'' degrees of freedom such as the center-of-mass (CM) and the ''relative'' (internal) degrees of freedom that exhibit non-existence of the total system's preferred structure \cite{LAP, ZurekStructures, Nelson}. However, in the context of LTS, such transformations may be allowed only if they are not in conflict with the actual local time distribution.

The following topics also require separate considerations. First, emergent (dynamical) nature of local time naturally raises a question of the possibly emergent {\it space}. In this context, ''space'' is naturally linked with the variables that determine the (sufficiently strong/weak) interactions as presented by the interaction terms $\hat H_{ij}$ in equation (\ref{eqn:4}) as well as with the (dynamical) formation of entanglement \cite{ramsdonk} in the system.
Second, relativistic causality as described by the universal speed of light in the vacuum should be reconsidered in the light of non-unique local times.
Finally, cosmological description of a transition from a dense early Universe to a present-day ''cold'' Universe consisting of the apparently decoupled (independent)  objects is needed \cite{Hartle}, together with a ''deeper'' physical foundations of the dynamical chain (\ref{eqn:3}), which we take as a primitive of our considerations. All those topics appear as  instances of the future testing of coherence and consistency of the Local Time Scheme and its implications.

\section{Conclusion}\label{sec:8}

Typically for the standard statistical and quantum physics, an ensemble description of large, many-particle systems is sufficient for most of the practical purposes. However, in the Local Time Scheme, such an description hides the physically rich and classically desirable description of the individual many-particle systems. Distinguishability, individuality and irreversibility are established for the {\it single}, exactly isolated many-particle systems--which consist of a large number of approximately isolated subsystems--in the spirit that is shared with the description of the classical ''macroscopic'' systems.

\section*{Competing interests.} We have no competing interests.

\section*{Author's contributions.} All authors equally contributed
to the paper.

\section*{Funding statement.} This research is funded by the Ministry of Education and Ministry of Science, Technological Development and Innovation, Republic of Serbia, Grants:
No. 451-03-65/2024-03/200124 and 451-03-65/2024-03/200122.

\section{Appendix A}\label{app:A}

Consider a pair $S=S_1+S_2$ of five-qubit and two-qubit systems with a time distribution $\{\mathcal{T}_1,\mathcal{T}_2\}$. Eigenergies are $E_{1}\in\{\pm5\hbar\omega_1/2,\pm3\hbar\omega_1/2,\pm \hbar\omega_1/2\}$ and $E_2\in\{0,\pm \hbar\omega_2\}$. Assume that $\langle\hat H_1\rangle=0$ and $\langle\hat H_2\rangle=0$, respectively. The parameters $\delta t_i$ may be used as the maximum values provided by the bounds $\tau_1=\pi/5\omega_1$ and $\tau_2=\pi/2\omega_2$. The choice of the mean energies admits the choice of the  equal probabilities in the initial state, $p=1/18$. Then, placing $p=1/18$ for every $\vert c_{ij}\vert^2$, instead of $\vert c_{ijk}\vert^2$ in equation (\ref{eqn:11}), the sum of the exponential terms obtained by all the  combinations of $E_{1i}\tau_1+E_{2j}\tau_2$ in the exponent give the sum:
\begin{equation}\label{eqn:A1}
\vert S(\mathcal{T}_1,\mathcal{T}_2)\vert^2 = \left\vert {1\over 18}\sum_{i,j} e^{-\imath(E_{1i}\tau_1+E_{2j}\tau_2)/\hbar}\right\vert^2 \approx 0.0292.
\end{equation}

\section{Appendix B}\label{app:B}

Consider the most general form of the four-system state (we omit the symbol of the tensor product):
\begin{equation}\label{eqn:B1}
\vert\Psi\rangle=\sum_{i,j,k,l} c_{ijkl} \vert i\rangle_1\vert j\rangle_2\vert k\rangle_3\vert l\rangle_4
\end{equation}

\noindent  where appear the subsystems' states constituting the respective orthonormalized bases, and the normalization condition $\sum_{i,j,k,l}\vert c_{ijkl}\vert^2=1$. Then the subsystems' states read, e.g.
\begin{equation}\label{eqn:B2}
\hat\rho_1=tr_{234}\vert\Psi\rangle\langle\Psi\vert=\sum_ip_1\vert i\rangle_1\langle i\vert, \quad \hat\rho_2=tr_{134}\vert\Psi\rangle\langle\Psi\vert=\sum_jq_j\vert j\rangle_2\langle j\vert,
\end{equation}

\noindent where $p_i\equiv\sum_{j,k,l}\vert c_{ijkl}\vert^2$ and $q_j\equiv\sum_{i,k,l}\vert c_{ijkl}\vert^2$, so that $\sum_ip_i=1=\sum_jq_j$.

Now, the first step (which we denote ''I'') in structuring the system is presented by equation (\ref{eqn:7}). Since merging the subsystems follows from the very strong local interaction, which dominates the merging-systems dynamics, we consider the instantaneous Schmidt form for both subsystems, $1+2$ and $3+4$; instantaneous Schmidt form is well defined even for the non-separable interaction between the subsystems \cite{MDScr2}. For the separable interaction between the local subsystems, the Schmidt form follows from the internal (strong-interaction induced) decoherence processes \cite{MDScr2,MDScr1}. Then equation (\ref{eqn:7}) reads
\begin{equation}\label{eqn:B3}
\vert\Psi(t_{12},t_{34})\rangle=\sum_{i,j,k,l,p,q} c_{ijkl} c^{ij}_{p}(t_{12}) c^{kl}_q(t_{34}) \vert p\rangle_1\vert p\rangle_2\vert q\rangle_3\vert q\rangle_4,
\end{equation}

\noindent with the same-form reduced states for both subsystems $1$ and $2$:
\begin{equation}\label{eqn:B4}
\hat\rho_{sI}(t_{12})=\sum_pr_p(t_{12})\vert p\rangle_s\langle p\vert ,s=1,2,
\end{equation}

\noindent where $r_p(t_{12}):= \sum_q\vert d_{pq}(t_{12},t_{34})\vert^2$ and $d_{pq}(t_{12},t_{34})=\sum_{i,j,k,l}c_{ijkl}c^{ij}_p(t_{12})c^{kl}_q(t_{34})$, while easy proved $\sum_{p,q}\vert d_{pq}(t_{12},t_{34})\vert^2 =1=\sum_pr_p(t_{12})$. Note that tracing out the subsystems ''wipes out'' the corresponding local times, e.g. as above: $\sum_q c^{kl}_q(t_{34}) c^{k'l'\ast}_q(t_{34})=\delta_{kk'}\delta_{ll'}$.

Now, the second phase presented by equation (\ref{eqn:10})
\begin{equation}\label{eqn:B5}
\vert\Psi(t_1,t_{23},t_4)\rangle = \sum_{p,q} d_{pq}(t_{12},t_{34}) \hat U_1(t_1)\vert p\rangle_1\otimes\hat U_{23}(t_{23})\left(\vert p\rangle_2\vert q\rangle_3\right)\otimes\hat U_4(t_4)\vert q\rangle_4;
\end{equation}

\noindent the time instants of the previous dynamics (local-time-distribution),  $t_{12}$ and $t_{34}$, are here the fixed parameters, not anymore the time instants.
On the use of instantaneous Schmidt form for the merging pair $2+3$, equation (\ref{eqn:B5}) takes the form of
\begin{equation}\label{eqn:B6}
\vert\Psi(t_1,t_{23},t_4)\rangle = \sum_{p,q,k} d_{pq}(t_{12},t_{34}) c^{pq}_k(t_{23}) \hat U_1(t_1)\vert p\rangle_1\otimes\vert k\rangle_2\otimes\vert k\rangle_3\otimes\hat U_4(t_4)\vert q\rangle_4,
\end{equation}

\noindent with the easy obtained reduced states:
\begin{equation}\label{eqn:B7}
\hat\rho_{1II}(t_1)=\hat U_1(t_1)\hat\rho_{1I}\hat U^{\dag}_1(t_1), \quad \hat\rho_{2II}(t_{23})=\sum_k\lambda_k(t_{23})\vert k\rangle_2\langle k\vert.
\end{equation}

\noindent In equation (\ref{eqn:B7}): $\lambda_k(t_{23}):=\sum_{p,q} \vert d_{pq}(t_{12},t_{34})\vert^2\cdot\vert c^{pq}_k(t_{23})\vert^2$, while $\sum_k\lambda_k(t_{23})=1$.

{}

\end{document}